\documentclass[preprints,article,accept,moreauthors,pdftex]{Definitions/mdpi}

\firstpage{1} 
\pubvolume{4}
\issuenum{1}
\articlenumber{0}
\pubyear{2022}
\copyrightyear{2022}
\usepackage{color}

\datereceived{02 April 2022} 
\dateaccepted{14 June 2022} 
\datepublished{} 
\hreflink{https://doi.org/}

\Title{Perturbative QCD Core of  Hadrons and  Color Transparency Phenomena}

\TitleCitation{Perturbative QCD Core of  Hadrons and  Color Transparency Phenomena}

\Author{Leonid Frankfurt $^{1,\dagger}$ and  Mark Strikman $^{2,}$*}

\AuthorNames{Leonid Frankfurt and Mark Strikman}
\AuthorCitation{Frankfurt, L.; Strikman, M.}

\address{%
$^{1}$ \quad  {Tel Aviv University;} leonidfrankfurt@gmail.com\\
$^{2}$ \quad {Penn State University}}

\corres{Correspondence: mxs43@psu.edu}

\firstnote{Current address: Tel Aviv University} 

\abstract{
In the current paper, we 
argue that the ground state of a hadron contains a significant 
perturbative quantum chromodynamics (pQCD) core as the result of color gauge 
invariance and of  the values of chiral and gluon vacuum condensates. 
The evaluation within the method of dispersion sum rules (DSR) of the vacuum matrix elements of the correlator of local currents  with the proper quantum numbers leads to the value of the radius of the pQCD core of a nucleon  of about  0.4--0.5 fm.
The selection of the initial and final states allows to select processes in which the pQCD core of the projectile gives the dominant contribution to the  process. 
 It is explained
that the transparency of nuclear matter for the propagation of a spatially small and color-neutral wave packet of quarks and gluons---a color transparency (CT)  phenomenon---for a group of hard processes off  nuclear targets can be derived in the form of the QCD factorization theorem accounting for the color screening phenomenon Based on the success of the method of DSR, we argue that a pQCD core in a hadron wave function 
is surrounded by the layer consisting of quarks 
interacting with quark and gluon condensates. As a result, in the quasi-elastic processes $e+A\to e'+N +(A-1)^{*}$, 
the
quasi-Feynman mechanism could be dominating  in a wide range 
of the momentum transfer squared,  
$Q^2$.    In this scenario,  a virtual photon is absorbed by  a single 
quark,  
which carries a large fraction of  the momentum of the 
nucleon and 
dominates in a wide range of $Q^2$.  CT should 
reveal itself in these processes at an extremely large $Q^2$ as the consequence of the presence of the  Sudakov form 
factors, 
which squeeze a nucleon. }


\begin{document}

\section{Introduction}

 The color transparency (CT) phenomenon is the suppression of the final and/or initial  state interaction 
for the small size wave packet of quarks and gluons produced in the hard processes  and propagated through 
a nucleus. The quantitative approach is to calculate the  cross-section of  the interaction of  small size wave 
packet  scattering off  a target, 
 as well as to describe some  properties of the  bound state hadrons specific for 
quantum chromodynamics (QCD).

CT has been derived (i) for the 
 {deep inelastic scattering (DIS)} 
processes initiated by highly virtual photons;
(ii)~for the processes of the diffractive electroproduction of vector mesons such as
$\gamma^*+A\to V+A$ for $V=\rho,\omega, \phi, J/\psi,\Upsilon, {\rm etc.}$; 
(iii) for the {processes:} $\pi+A\to {\rm two \,\, jets} +A$,
$p+A\to {\rm three \,\, jets}+A$; 
(iv) in the case of small 
parton momentum fraction, small-$x$ processes, factorization theorems can 
be derived for the diffractive photoproduction of the
bound states of heavy quarks   such as 
$J/\psi$ 
 and 
$\Upsilon$, produced
in the ultraperipheral collisions of heavy ions. 
and
(v) CT is generalized to include effects of the leading twist gluon 
shadowing for small $x$ processes.

Within QCD, a hadron consists of the three overlapping   layers, 
corresponding to two distinctive phases of the QCD matter. The outer  layer is formed  by the pion 
cloud of a hadron. Just this layer produces internucleon attraction in low-energy nuclear phenomena. The next layer  is 
formed by quarks interacting with chiral and gluon condensates.  The existence of this layer is the basis of the success of 
the method of dispersion sum rules in the calculation of 
parameters of the ground states of hadrons. The important role of the vacuum condensate of chiral quark pairs is 
implied by  the phenomenon of spontaneously broken chiral symmetry in QCD. The boundary between both layers can
 be evaluated as the boundary of the region where two-pion exchange between nucleons dominates.
This value of the boundary was estimated in
Ref.~\cite{Drell} 
long ago. In any case, the position  of this  boundary is not  well defined  since it fluctuates.

 The second layer is relevant for the competition between soft and hard processes for the large momentum transfer 
behavior of hadron form factors. This competition  results from the specific of the Lorentz transformation within 
light cone quantum mechanics.  It was  found that for a two-body system that large momentum transfer $Q$ is multiplied
by the factor, 
$1-\alpha$, 
in the argument of the wave function of a final state of a two-body system~\cite{Kogut:1972di}. 
A similar pattern holds for many body systems. Thus, the relative contribution into form factors of configurations where one constituent carries most of the 
 light cone (LC) 
fraction---$\alpha$---of the hadron is enhanced. Feynman has pointed out that  this 
may lead to a dominance of the configurations where one parton carries practically all the momentum of the hadron. Hence,
 this contribution is referred to as the Feynman mechanism. In practice, there is a wide pre-asymptotic region, 
 e.g.,  $\alpha \geq 0.8 $, which is strongly enhanced in  a wide 
range 
of $Q^2$.  
In what follows, this 
kinematics is referred to 
as a quasi-Feynman mechanism.

  Together, these two layers describe the QCD phase of the spontaneously broken chiral symmetry.
The significant probability of the 
perturbative QCD (pQCD)
core within the wave function of a hadron follows from the analysis of the vacuum 
correlator of local currents in the coordinate space,  i.e., effectively from the color gauge invariance and the values of the 
chiral and gluon condensates. Thus, the evaluation of the radius of the pQCD core of a hadron requires using a model 
accounting for the condensates.

 The discovery of heavy quarks such as  $c, b, {\rm etc.}$ 
allows to expand CT to the number of the processes, where CT has 
already  been observed and can be further investigated.

 The 
understanding of  the 
QCD structure of  the wave function of a hadron and  of
a nucleus allows to separate a group of  hard processes, where 
the hard interaction produces spatially small wave packets of quarks and gluons.  The strength of  the  interaction  of such 
wave packets  with hadrons is  unambiguously calculable within QCD for high-energy processes. Selecting special initial and final states  is necessary  for ensuring  the dominance of the contribution of  the pQCD core of the  hadrons. For  this group of phenomena, CT is  one of the elements 
 of the  QCD factorization theorem. In these processes, 
CT was already unambiguously observed.  The second group is formed by the
 processes,
where there is no constraint for
 the pQCD core to dominate in the wave function of a hadron. In this case, the
 quasi-Feynman mechanism dominates in a wide range of momentum transfer.  At an extremely large momentum transfer, the contribution of the pQCD core should dominate since the Sudakov form factors gradually squeeze the  wave packet.
 
\section{Three-Layer Structure of the  Nucleon Wave Function in QCD}

\subsection{The Spatial Distribution of Valence Quarks in  a Nucleon}

 The valence quark and momentum sum rules for the parton distributions within a hadron unambiguously follow 
from the Wilson operator expansion. For certainty, 
the 
discussion in this paper is restricted to  
the case of a nucleon target.
 The sum rules for the generalized  
valence quark distributions, $V_N(x,Q^2,t)$,
 at a non-zero momentum transfer,
\begin{equation}
 \int V_N(x,Q^2,t)dx=F^{V}(t), 
\end{equation}
 is of a prime interest to us.
Here, $F_N^V (t)$ is the isotopic (SU(3)) vector form factor of a nucleon calculable in terms of the combination of the electromagnetic form factors of a proton and a neutron. The generalized  valence quark sum rules follow  for  any 
$Q^2$ from the combination of the Ward identities for electroweak currents  and the energy dependence of the high-energy amplitudes in QCD in  non-vacuum quantum numbers in the crossed channel~\cite{Alfaro}.  To investigate the pQCD core of a nucleon at moderate $Q^2$, 
the method of dispersion sum rules (DSR), developed  
in~\cite{Shifman:1978bx}, is used here.

 The radius of valence quark distribution follows from the sum rules and from the data on the electromagnetic form 
factors of a nucleon: 
\begin{equation}
(r_V^2)^{1/2}=0.65\, {\rm fm}. 
\end{equation}

\subsection{The pQCD Core of the Wave Function of a Nucleon }

 The  Dirac sea is an  important property of  relativistic quantum field theories. Ignoring  the Dirac sea in the 
non-relativistic approximation leads to the non-conservation of baryon and electric charges of the energy---momentum tensor 
as defined in QCD and a related violation of probability conservation for the high-energy processes.  For example, these violations lead to the  so-called West correction for  the structure functions of the  
 deuteron,
$\sigma_{\rm tot} 
(eD) < \sigma_{\rm tot} (ep) + \sigma_{\rm tot} (eN) $ in the impulse 
approximation~\cite{West:1971oyz}.
This  calculation used the non-relativistic wave function of the deuteron. Such a correction  results in  the
 violation of the exact QCD sum rules such as the baryon charge sum rule, in the violation of the Glauber decomposition over rescatterings, etc.

    The only  approach   known so far to account  for the Dirac sea in a way consistent with the exact QCD sum rules is 
to use  the 
 LC. 
mechanics of  nuclei.   It  resembles  the parton model approximation  for  a quantum field 
theory,
 suggested by Feynman. Imposing angular momentum conservation and the requirement  of Lorentz symmetry for on-mass-shell 
amplitude, 
LC mechanics can be transformed into the  instant time form for a wide range of nucleon momenta~\cite{Frankfurt:1981mk}. In the LC mechanics of a nucleus, the West correction disappears, as can be 
found 
in Ref.~\cite{Frankfurt:1981mk}. 
 This phenomenon is 
important in nuclear theory, in the 
kinematics, where 
 $k^2/m_{N}^2$ 
is not negligible; here, $k$ denotes nucleon momentum  and $m_N$ denotes nucleon mass.  
 
 The phenomenon 
is 
also present   in quantum electrodynamics (QED)  
in the calculation of high-order corrections to the wave functions of molecules.

 To investigate the properties of an LC wave function of a nucleon, 
the vacuum element of the retarded commutator 
of local color-neutral currents, $J$, with the quantum numbers of a 
nucleon,  
 Omitting Lorentz indices we can write 
\begin{equation}
K(y_0, \vec{y})=\left < 0\right| \theta(y_0)  \left[J(y_0,\vec{y}), 
J(0)\right]  \left |0\right>. 
\end{equation}
 are analyzed here. $\left < 0| \right.$ and $ \left. | 0\right>$ denote in and out states,  $y$ is the difference of 4 D coordinates of the in and out currents, and $y_0$ is the zero component of $y$.     $\theta$ is the Heaviside function,  

The  
retarded commutator is equivalent to the $T$- product but allows the analytic continuation of the 
Fourier transform into the complex plane of energies  and  derives dispersion relations.
This correlator in the momentum space was analyzed in 
Refs.~\cite{Shifman:1978bx, Ioffe:2005ym}  within  the DSR 
approximation. For the aims of the study, given here,
it is convenient to analyze this correlator in the coordinate~space.

The intermediate states in the correlator with the quantum numbers of a nucleon  are accounted for as the full system  
of eigen-states of the QCD Hamiltonian: 
 \begin{equation}
K(y_0, \vec{y}))=\sum_{n} \left < 0 \right | J(0) \left | n \right > ^2 \exp(i(E_{n}-E_0)y_{0}-i(p_{n}\cdot \vec{y}) )\theta(y_0)d \tau_{n}.
\label{Correlator}
\end{equation}
 Here, $E_n$ and $E_0$ are ..., respectively; $p_n$ is ..., and $\tau_n$ is ... 

In the Euclidian domain, in the limit 
$i(E_0-E_n)y_0\to \infty$, only the contribution of  the state with the minimal mass (\emph{n} = 0)  survives. 
To improve the convergence with the  DSR 
 approach, one subtracts  the contribution of the intermediate states with the masses 
significantly larger than the nucleon mass. The subtracted contribution is calculable in pQCD~\cite{Ioffe:2005ym}.

 The initial condition for the space--time evolution for the valence quark component of the wave function of a 
hadron is 
\begin{equation}
\phi_{N}(t=0,\vec{y}=0)=\left<0\right|J(0)\left | N\right>/(E_N-E_0).
\label{}
\end{equation}

The current, $J$, is local in the coordinate space to ensure the conservation of the color gauge invariance. At the 
starting 
point of  the  space--time evolution, the pQCD wave packet  has the zero size. The hierarchy of the initial conditions is 
regulated by the asymptotic freedom combined with the approximate conformal invariance of QCD. In the 
 actual
calculations, 
 one choses 
the current, $J$, 
following 
 Ref.~\cite{Ioffe:2005ym}, 
where
the spatially local and color-neutral current $J=J^{3q}$
was used.
 The currents, containing a larger number of quark--gluon fields, were used for evaluating the 
 gluon distribution  in the  nucleon; 
 see
discussion in Ref.~\cite{Kolesnichenko:1984dj}.

 The investigation of the space--time evolution of the  nucleon wave function with the increase in the relative distance 
$y$ between constituents allows to identify the coexistence of the three distinctive layers in the wave function.

 The  radius  of the pQCD core of  a nucleon can be estimated based on  the analysis of the  correlator, $K$, in the coordinate 
space performed in Ref.~\cite{Shuryak:1993kg}. The analysis uses the standard values of the quark and gluon 
condensates and 
indicates that the correlator is close to the free 
 correlator 
for the radius 
$r \le r_c$, where  
\begin{equation}
r_c \approx (0.4 -  0.5)\, 
 {\rm fm}.  
\label{r_c}
\end{equation}

 The pQCD core of the radius $r_c$    is surrounded by a layer of nonperturbative QCD phase where  bare quarks 
 and antiquarks interact with the vacuum condensates and obtain masses. As a result, the properties of this layer resemble the 
constituent quark model of a nucleon. The thickness, $\Delta _{\rm n.p.}$,  
of the nonperturbative layer 
can be estimated from 
\begin{equation}
\left <  r \right>_{N}=\Delta_{\rm n.p.} +r_c, 
\end{equation} where
$\left <r \right>_{N}$  is the radius of a nucleon. For 
illustration, let us  
 choose 
$\left<r\right>=0.85$~ fm. Thus, 
$\Delta_{\rm n.p.}\approx (0.35$--
0.45)~fm. 
 It is natural to separate the pion cloud of a nucleon from the  nonperturbative layer.
$\Delta_{\pi}\approx 0.2$~fm is the thickness of the  pion cloud calculated from the contribution of the two-pion state in the 
electromagnetic form factors of a nucleon~\cite{Drell}. Thus, $\Delta_{\rm n.p.}-\Delta_{\pi}\approx 0.2$~\rm{fm.}

 To conclude, the derived form of the nucleon wave function follows in the weak coupling regime of QCD from the local 
color gauge invariance, asymptotic freedom, and from the small density of the vacuum condensate of the chiral pairs. 

\subsection{Lattice QCD as the Tool to Probe Nucleon Wave Function}

 The lattice evaluation of vacuum correlators $K$ (\ref{Correlator}) may help improve  the evaluation of the
nonperturbative QCD phenomena beyond the mean field approximation used in this paper.  

The  three-layer structure of a nucleon reveals itself in the distinctive QCD phenomena:
 
(i)  The generation of the running masses of constituents due to their interaction with vacuum condensates was discussed
in Ref.~\cite{Diakonov:1995zi} but without a pQCD core. As a result, the second layer resembles the  constituent quark model. 
The educated guess for the thickness  of this phase is between $0.4$ fm and $0.6$ fm.

 (ii)  The investigation of the transitions between the nonperturbative  and pQCD phase  would provide  a new 
probe of the role of the spontaneous violation of the chiral symmetry in the structure of the  ground states of 
 hadrons.

\subsection{The Proof of the Presence of pQCD Core within a Hadron}

 The very existence of the pQCD phase within a nucleon  differes ffrom the expectations of 
the nonrelativistic nuclear theory 
at the scale comparable to the scale observed in the low-energy nuclear phenomena. Thus,  
an  additional theoretical reasoning is presented below.

 The existence of the pQCD core  follows directly from the QCD factorization theorems derived in the momentum 
space representation. Let us consider the  example  of the process involving a pion.

(i) The dominant high-momentum quark-antiquark, $q\bar q$, 
component of the pion wave function is given 
in QCD by the one gluon exchange:
\begin{equation}
\psi_{\pi}(k) =\alpha_s f_{\pi} C_{\pi}/k^2.          
\label{zerowf1}
\end{equation}
Here,  $f_{\pi}$ is the constant, determined from the pion $\beta$ decay,  
$C_{\pi}$ is a constant calculable in pQCD,
$\alpha_s$ is the running coupling constant, and
$k$ is the momentum of quark (antiquark). The product $f_{\pi} C_{\pi}$ is calculable in the approximation of the partial conservation of the axial current.  The Fourier transform of this wave function into coordinate space gives
$\psi_{\pi}(r=0)$.  The high momentum tail of the wave function of pion in the momentum space has been measured at 
FNAL  (Fermi National Accelerator Laboratory, Batavia, IL, USA)
in the experiment, where
CT was also discovered
in the  reaction $\pi+A\to 
{\rm two\,\, jets} 
+A$~\cite{E791:2000kym}
with the pattern, predicted in Ref.~\cite{Frankfurt:1993it}. A similar behavior is expected for the scattering 
of any 
meson. The only difference is the dependence of the probability of the tail on the type of a~meson.

  (ii) The factorization theorem of QCD for two-body processes see 
Equation~(\ref{zerowf1}) is expressed through the wave 
function of the quark--gluon core of a pion. The existence of the quark--gluon core of a nucleon unambiguously follows from the 
QCD factorization theorem. The direct experimental observation of the quark--gluon core of a nucleon 
in the process $p+A\to {\rm three\,\, jets} 
+A$  is  much more difficult than that for the pion since the pQCD contribution into a hard process  is significantly 
smaller---$\propto \alpha^2_s$. Additionally,  the absolute value of the high-momentum tail of a parton wave function of a nucleon 
is the subject of modeling.   

\subsection{Pion Field in QCD}
 Internucleon attraction in low-energy phenomena is mostly due to the interaction of the pion clouds of the interacting 
nucleons. The pion clouds are located at the nucleon outer  layer.  Internucleon repulsion is dominated by the 
nucleon core. The long-range pion field of a nucleon has been derived within QCD from the phenomenon of a spontaneously 
broken chiral symmetry. The pion mass was calculated in QCD in terms of the chiral vacuum condensate,  
$\left<0\right |q\bar q\left |0\right >$~\cite{Gell-Mann:1968hlm}. The pion mass arises due to the explicit violation of chiral 
symmetry by the term in the Lagrangian containing non-zero masses of up ($u$) and down ($d$) quarks~\cite{Gell-Mann:1968hlm}:
\begin{equation}
m^2_{\pi}=(m_u+m_d){ \left<0\right |q\bar q\left |0\right >\over f_{\pi}^2}.
\label{GOR}
\end{equation}

 The concept of the pion cloud of a nucleon is valid within the restricted kinematical region where a virtual pion is close 
to the pion mass shell  because a pion is  a collective mode.  The experimental restriction is that the antiquark distribution 
in a nucleon is rapidly decreasing in a nucleon  as
 $\bar q(x,Q^2)\propto (1-x)^{m}$.  %
Global parton distribution fits typically give $n(Q^2)\approx 7$. 
Thus, 
the distribution of 
 antiquarks mostly originate from  the second layer where antiquarks obtain masses as a result 
of the interaction with the chiral condensate. The role of the second layer, for the antiquark distribution within a nucleon is 
 decreasing with increase in nuclear density. This pattern is the opposite to the expectation of the models where pion 
 condensate is present in the cores of neutron stars. Within QCD, the pion field in a hadron is reduced with a significant decrease in the overall size of the hadron due to effects of color screening and color gauge invariance. 
This phenomenon is the property of the weak coupling regime where the interaction is decreasing with the decrease 
of 
 the overall size of a hadron{;  
see} discussion below.
This QCD phenomenon is manifested, for example, in the  decays of the excited state of charmonium and bottonium 
and may be tested  by lattice QCD methods. This phenomenon is opposite to the experience of the low-energy 
nuclear physics and it is absent in the bag models. This is because the bag models of a hadron---in contrast to the
full QCD---do not include  color screening  and asymptotic freedom phenomena. Therefore, in the bag models, a position of the hadron bag surface does not depend on the distribution of quarks 
and gluons. This feature also results in the violation of  causality. Thus, QCD implies that the pion field is not a universal 
property of a hadron. One of the striking experimental confirmations was the discovery that for $J/\psi$ and other onium 
mesons, interaction with pion field is suppressed.

 Thus, the preQCD models of nucleon--pion interactions differ from QCD. This difference  reveals itself for the large pion 
momenta, as can be  found
in  Ref.~\cite{Koepf:1995yh} (and 
references  
therein). For the distances within a nucleon that are significantly smaller than the 
thickness of a pion cloud, different degrees of freedom dominate within QCD.

 Thus, the difference between the preQCD models of nucleon--pion interactions and the expectations of QCD reveals itself 
for the large pion momenta and large pion virtualities. For the 
distance scale  within a nucleon that is significantly smaller than the thickness of the pion cloud, different degrees of freedom 
dominate in QCD. The found properties of the wave function  of a nucleon are sufficient to resolve the zero charge 
puzzle: in a preQCD quantum field theory,
~ultraviolet divergencies nullify the interaction of hadrons~\cite{Landau}. In QCD, 
ultraviolet divergencies  present in the preQCD quantum field theories disappear due to the composite structure of a hadron.
the wave function of a hadron contains in its central region  the  quark--gluon pQCD core instead of meson fields characteristic for preQCD field theories as the consequence of the color gauge invariance, asymptotic freedom,  and the small density of vacuum chiral condensates.  The scale where the nonperturbative QCD surface effects are transformed into  the 
 quark--gluon core of a hadron plays the role of cutoff within  preQCD field models within 
 effective field theory (EFT). 
 
Let us note that the  related phenomena in hot nuclear matter are beyond  the scope of this paper.

 \section{The Properties of QCD Which Lead  to CT Phenomena}
 
 The wave packet,  produced in the hard processes, is built of highly  virtual quarks and gluons, which weakly 
interact with the nuclear environment~\cite{Mueller:198t2bq}. 

 In Ref~\cite{Brodsky}, it was assumed that in the elastic $\pi N $ large angle collisions, the pion and nucleon are 
squeezed 
leading to the cross-section of a quasi-elastic reaction off nuclei $\pi A\to \pi N (A-1)^* $ equal to $A$ times the elementary cross-section. There are experimental indications for a moderate  increase in the transparency 
of nuclear matter with an increase in $Q^2$ for the  electroproduction of pions and $\rho$ mesons, as can be 
 found in 
review
\cite{Dutta:2012ii}.


 The experiments, designed to search for the CT in the quasi-elastic processes, $e+A\to e'+N+(A-1)$, did not observe 
an increase in transparency up to $Q^2=14\, \mbox{GeV}^2$ \cite{HallC:2020ijh} . 
This result is qualitatively consistent with the analyses of  
the data on electromagnetic form factors  at the intermediate $Q^2$ which indicate that the dominant contribution to the nucleon
~\cite{Ioffe:2005ym} and pion~\cite{Radyushkin:2004sr} form factors  is the electron scattering off configurations in which 
 the virtual photon, 
$\gamma^*$, 
is absorbed  by  a leading quark which gained its momentum distribution in the interaction with vacuum condensates of quarks and 
gluons. Thus, the data are inconsistent with the democratic chain approximation, suggested in {Ref.Ref.~\cite{Brodsky:1974vy} 
at 
achievable $Q^2$.

\subsection{The Conditions for  the Validity of CT}

 The direct observation of the CT phenomena in the hard high-energy processes  requires the validity of two conditions:

(a)  The interaction between hadrons depends on the value of the region occupied by color within the interacting 
hadrons. This feature of QCD is  proved for hard processes in the form of QCD evolution equations.
If the  transverse component of the radius of a hard projectile, $r_t$, perpendicular 
to the wave packet momentum is sufficiently small,  the forward amplitudes 
are proportional to 
$r_t^2$:
\begin{equation}
\sigma_{\rm tot}(a+b)=c r_t^2, 
\label{CS}
\end{equation}
 where $c$ is the speed of light.

 Equation~(\ref{CS}) is another form of the  Bjorken scaling for the total cross-sections of DIS  which follows from the 
approximate conformal invariance of QCD~\cite{Frankfurt:1988nt}. Equation~(\ref{CS}) has been confirmed in many high-energy experiments.
In particular, it serves as a basis for the description of exclusive and inclusive experiments  and can be formulated in the form 
of QCD factorization theorems for these processes. Note that the account of the QCD evolution in $x$ and $Q^2$  leads to the somewhat weaker dependence on $r_t$.  However, the physics relevant for this extreme kinematics is beyond the scope of 
this paper. Thus, the first condition of the validity of CT is the presence of the trigger, which selects a squeezed quark--gluon 
configuration within a hadron, and a part of the proof is the derivation of the dipole model~\cite{Blaettel:1993rd, 
Frankfurt:1993it,Frankfurt:1996ri}. The spatial size of the produced wave packet of quarks and gluons should be 
significantly smaller than the radius of a hadron.

 (b) 
The second requirement is  that the lifetime, $L_c$, i.e., the distance propagated by sufficiently energetic but squeezed 
quark--gluon wave 
packet,
should be sufficiently large  to avoid expansion while traversing  the nucleus. Thus, 
Einstein time dilation  plays an important role in the CT. The coherence length was derived  from the theoretical analyses  
of the Fourier transform of the structure functions of a target nucleus/nucleon  in coordinate space. In the rest 
frame of the nuclear target, at large $Q^2$~\cite{Gribov:1965hf,Ioffe:1969kf}:
\begin{equation}
z=L_{c}=(1/2m_Nx).
\label{Lc}
\end{equation} 
Here, $Q^2\approx (m_q^2+k_t^2)/z(1-z)$ is the resolution of the investigated configuration, 
$m_q$ is  the quark mass. 
$z$. 
is a fraction of photon momentum
carried by the produced quark, and $k_t$ is its transverse momentum. This equation follows from the uncertainty 
principle for the transition $\gamma^*\to q\bar q$ and Einstein time dilation. 
Numerical calculations show that Equation~(\ref{Lc}) is also valid within the parton model for moderately large  $x$. Thus, 
the  pQCD core 
for a longitudinally polarized photon together with the color-screening phenomenon 
guarantee  the validity of CT if the lifetime for a wave packet describing the pQCD core
is sufficiently large to traverse a~nucleus. 

\subsection{Discovery of CT}

 The search for the  CT  allowed to establish at what resolution scale $Q^2$  the basic degrees of freedom 
characterizing the wave function of a nucleon of a nucleus are   quarks and gluons. This  would allow to determine 
an upper bound for the range of virtualities for which EFT approaches  can be used.

 The discovery of a small cross-section of a diffractive photo-production of $J/\psi$  meson was the first observation 
of CT.   The value of the $J/\psi$--$N$  cross- section,  extracted within the  vector dominance model (VDM) 
of  $\sigma(J/\psi$--$N)\approx 1$~mb, was much smaller  than 
the genuine cross-section 
 $\sigma_{\rm tot} (J/\psi$--$N) \approx 4$~mb, 
extracted from the $A$-dependence of  
$J/\psi$ 
quasi-elastic photo-production.  The difference originates from the production of 
$J/\psi$ in a small size configuration, whose size is 
 about 
 $1/m_c$~\cite{Frankfurt:1985cv}, where $m_c$ is the c-quark  mass. 
 The phenomenon is 
especially striking when one compares 
 $J/\psi$--$N$ and $\psi'$--$N$
cross-sections, extracted from the data using VDM.   VDM leads to 
$\sigma _{\rm tot}(\psi'$--$N) <  \sigma _{\rm tot}(\psi$--$N)$, while  in QCD, the opposite trend is expected since the transverse 
size of $\psi'$ is exceeds twice 
that of $J/\psi$.

 CT  is an unambiguous prediction of QCD for the phenomena, where QCD factorization theorems  are applicable:
 
(i)  DIS processes off a nuclear target:
\begin{equation}
 \sigma_{\rm inel}(\gamma^*_{L}+A)\to A\sigma_{\rm inel}(\gamma^*_{L}+N),
\end{equation}
 where $\sigma_{inel}$ denotes the inelastic cross-section, and $\gamma^*_{L}$ stands  for the longitudinally polarized  photon. 

At small $x$, CT is valid 
 as soon as 
the leading twist nuclear  shadowing is taken into account~\cite{Frankfurt:1988nt}.  

 (ii) The cross-section of the diffractive electroproduction of longitudinally polarized vector meson off  nucleon and nuclear 
targets has been predicted in Refs.~\cite{Brodsky:1994kf, Collins:1996fb}. The derived QCD factorization theorem  is also applicable 
 to 
a 
nuclear target. For the invariant momentum transfer,  $t=0$, and the large $Q^2$ cross-section of the diffractive electroproduction 
of the vector meson, $V=\rho, \omega, \phi $, 
CT 
unambiguously follows from the QCD factorization theorem~\cite{Collins:1996fb}.

 (iii) In the case of the diffractive photo-production of a heavy quarkonium, a heavy mass of $c$ and 
$b$ quarks  guarantees the 
applicability of the 
QCD factorization theorem. These processes were observed 
at HERA facility (at German Electron Synchrotron DESY, Hamburg, Germany) 
and 
in 
ultraperipheral heavy-ion collisions 
at the  LHC (Large Hadron Collider at the European Organization for Nuclear Research CERN, Geneva, 
Switzerland).

For $t=0$, 
\begin{equation}
d\sigma(\gamma+A\to V + A)/dt =A^2 d\sigma(\gamma+N\to V  +N)/dt,
\label{shadvm}
\end{equation}
where
$V=J/\psi$ or 
$\Upsilon$.
  
 One can employ the observation from quarkonium models that the radii of the onium states 
are  much smaller than 
those for light vector mesons and pions: $r_{J_/\psi} =0.2$~ fm  and $r_{\Upsilon } =0.1$~fm. 
Note here that for a small $x$, expression (\ref{shadvm}) has an additional factor---the square of the ratio of gluon 
densities  in the  nucleus (per nucleon)  and in the~nucleon. 

 (iv)   Diffractive production by the pion projectile of two  jets off a nuclear target, $ \pi+A\to {\rm two\,\,  jets} +~ A$. The 
process
 was observed at FNAL~\cite{E791:2000kym} and predictions~\cite{Frankfurt:1993it} for the longitudinal and transverse 
momentum of jets and for the $A$-dependence of the cross-section  were confirmed.
The same effect is expected for kaon beams and for the fragmentation of a proton into three jets.

The restriction on the region of applicability of CT, given by Equation~(\ref{Lc}), can be somewhat weakened by accounting for the 
phenomenon of quantum diffusion in the space--time evolution of small size wave packet~\cite{Frankfurt:1988nt,Farrar:1988me}: 
\begin{equation}
r^2_{t}(l) =\left[r_{t}(0)^2 + \left(\left<r^2\right > - r_{t}(0)^2\right)(l/L_c)\right ] \theta(L_c-l)+\left<r^2_t\right>\theta(l-L_c).
\label{QD} 
\end{equation}
\noindent 
Here,  $r_t(l)$ is the transverse size of the wave packet which evolves with distance $l$ from the interaction  point. Relativistic kinematics is considered where
$l$ is equal to the distance from the interaction point);
 $\left<r^2_t\right >$ is a nonperturbative transverse radius squared  of the wave packet. The region of applicability of the above
  formulae is that $\sqrt{r^2_t}$ should be significantly smaller than the radius of a hadron.

\subsection{On the Quasi
Feynman Mechanism in Quasi-Elastic  Processes at Achievable $Q^2$ }

 In the relativistic theory, due to specifics of the Lorentz transformation, the quasi-Feynman mechanism 
may dominate in the nucleon and meson form factors  up to large $Q^2$ (which are likely to grow with the increase in 
the number of  constituents in a hadron).  In this mechanism, the dominant contribution to the form factor  originates  from the 
configurations, in which one parton carries nearly all the LC  fraction of the momentum of the nucleon.   In these configurations, 
squeezing is moderate while  the correlation length in a wide range of $Q^2$ is comparable to the internucleon distance.  
As a result, the  expected CT effect is rather mild up to large $Q^2$.

 To visualize a relativistic description, let us 
choose 
the transferred momentum, $q_t$, to be 
 orthogonal   to the quantization axis. This condition can be achieved by choosing 
the center-of-mass 
of the $e N$ system and
 setting
the electron momentum, $P\to \infty$~\cite{Drell:1969ca}.  The analysis of the energy denominators of 
the
LC  mechanics  shows that 
$q_t$ is multiplied in the argument of the wave function of the final nucleon by the 
factor 
$(1-\alpha)$ ~\cite{Kogut:1972di}.. Here, $\alpha$ is the fraction
of a nucleon momentum, carried by the interacting constituent. 

Let us consider 
the two-body system 
\cite{Kogut:1972di}:
\begin{equation}
F(q_t^2)=\int \psi\left({m^2+k_t^2\over \alpha(1-\alpha)} - 4 m^2\right)\,
\psi\left({m^2+[k_t+q_t(1-\alpha)]^2 \over \alpha(1-\alpha)}-4m^2\right) S^2(q_t^2/k_t^2)\:\:  d\alpha d^2k_t \\
\end{equation}
Here, $\psi$ is the wave function of two body system $m$ denotes the constituent  mass, 
$S(Q^2_t/k_t^2)$ is the Sudakov form factor, which squeezes the wave 
function of a nucleon, and 
the 
production of 
$N\overline{N}$ 
pairs from the vacuum is neglected.
The  coherence length (\ref{Lc})
accounts for the space--time evolution of the produced wave package through an uncertainty principle. 
A
different definition of the coherence length was used in Ref.~\cite{Miller}.

The value of squeezing depends on the  minimal $k_t$ when radiation is allowed.  
Here,
$k_t^2\approx (\Lambda_{QCD})^2$ is used
as it is popular
in the description of hard processes~\cite{Dokshitzer:1978hw}.  
Thus, at $Q^2_t\to \infty$, the two-body system is squeezed.  For the realistic $q_t$, maximal contribution arises from 
the region, where $\alpha$ is 
quite 
close to one.  This resembles  the Feynman mechanism where the leading parton 
carries the whole momentum of a hadron. In the mean field approximation, the wave function of a nucleon is effectively modeled 
as a two-dimensional harmonic oscillator,~as can be 
 found
in Ref.
\cite{Miller} 
(and references
therein). Thus, in these models,  the  quasi-Feynman mechanism dominates  in a wide range of the
momentum transfer (actually 
 for
any $Q^2$) 
and 
 can 
explain~\cite{Miller} the lack of CT effect, reported in the 
JLab (Jefferson Laboratory, Newport News, VA, USA) 
experiment~\cite{HallC:2020ijh}. In a general case, a wave function contains the short-range  correlation between quarks. 
This correlation is moderately squeezed by the Sudakov form factor accounting for the lack of radiation in the 
process considered. 

 Let us analyze  the $e+A\to e'+p+(A-1)$ process. The kinematical restriction is   $x=Q^2/2q_0m_N=1$, 
where $q_0$ the virtual photon energy.  
Let
$M$ 
 be 
the 
mass of a state within a nucleon corresponding to the Feynman mechanism. To evaluate $M$ for the 
$\left|3q\right >$ (three-quarks) 
system, 
the average transverse momentum of a leading parton within a nucleon  is set to 
$k_t=0.3(0.4)$~GeV $/c$. Thus, 
\begin{equation}
M^2\approx q_t^2(1-\alpha)/\alpha+\sum_{i} k_t^2/\alpha_i \approx {q_t^2(1-\alpha)\over \alpha}+ {4k_t^2\over 1- \alpha}.
\end{equation}
Here,
the running masses of quarks are ignored in order 
to simplify the description and also consider $q\bar q$ as a state rather than a three-quark   state,
approximating two recoil quarks by one massless parton (accounting for  a finite mass of the recoil, 
two quarks would further increase $M$). For a fixed $k_t^2, Q^2$, the minimal value of $M^2$ is reached 
for $(1-\alpha)/\alpha=\sqrt{4k_t^2/Q^2}$. Thus, a produced state should have a minimal mass squared around  
\begin{equation}
M^2\ge 2 \sqrt{Q^2 4k_t^2}
\label{mass}
\end{equation}

For $Q^2=7 (14)$~GeV$^2$,  one finds $M^2\approx 1.6$ to 
2.1  (2 to 
3) GeV$^2$. 

The
 coherent length 
\begin{equation}
L_{c}=2q_0/(M^2-m^{2}_N)=
\frac{Q^2 x}{m_{N}(M^2-m_N^2)}.
\label{lc1}
\end {equation}

Thus, for $Q^2=14 $~GeV$^2$, Equation~(\ref{lc1}) leads to  $L_{c}\approx 2$ to 
3~fm. $L_c$ may appear even smaller since here it is 
ignored that gluons carry a fraction of the nucleon momentum, as well as the finite mass of the recoil two -quark state.
Thus, the produced wave packet rapidly expands to the  nucleon size. Note also that due to the presence of the growth of 
$M^2 \propto \sqrt{Q^2}$ with increase in $Q^2$, the increase in $L_c$ is slower than found in Ref~\cite{Farrar:1988me}. This 
corresponds to a  relatively slow transition to the hard regime accompanied by a CT phenomenon.  
Experiments 
at TJNAF (Thomas Jefferson National Accelerator Facility, JLab) 
did not find a signal of CT in the 
quasi-elastic processes $e+A\to e'+p+(A-1)$ 
 at 
$Q^2=14$~GeV$^2$ \cite{HallC:2020ijh}
which may be due to both 
the slow increase in squeezing and 
a small and slow increase 
with the energy   correlation length.
Originally, in Ref. 
 \cite{Farrar:1988me}, the coherence length was estimated using a nonrelativistic 
quark model, leading to $\Delta M^2 = M^2 - m_N^2 \sim 0.6$~GeV$^2$. 
Taking $M \sim 2$~GeV, based on the mass spectrum of baryons with the nucleon quantum numbers, 
would reduce $L_c$ by a factor of five and completely wipe out the 
CT effect at 14 GeV$^2$. Hence, the  current 14 GeV$^2$ data cannot a priori distinguish a short  
coherence length scenario and the Feynman mechanism scenario.

Also  it is worth  noting  that
further analysis of the data~\cite{HallC:2020ijh} with realistic spectral functions is necessary since the 
transparency for scattering off carbon differs by 40\%  for the 
$s$- and $p$-shells~\cite{Frankfurt:1990pz},  
 while the short-range 
correlations, which constitute  15 to 20\%
of the spectral function, are not included at all. In addition, the EMC (European Muon Collaboration, CERN) 
effect 
leads to a small but non-negligible reduction in transparency, somewhat masking the increase in transparency due
to CT~\cite{Frankfurt:1994nn}.

 A promising way to investigate the Feynman mechanism as the mechanism of the onset of CT is to investigate the high-$Q^2$ process 
$e+d\to e+p+n$ in the kinematics, where the effective distances between nucleons in the deuteron are at approximately 
1--1.5 fm.  
This 
can be achieved by choosing the momentum of the nucleon spectator to be approximately 200~MeV/$c$ and to investigate the dependence of 
the cross-section on $Q^2$ at $x\approx 1$.  The cross-section has a minimum and maximum as a function of the momentum of the nucleon 
spectator, which can be used as the effective tool to search for CT~\cite{Frankfurt:1994kt}.

We argued above that the pion field is presumably already suppressed in the hard processes  
 \hl{at} 
 $Q^2$ of about few GeV$^2$.  
This should result 
 into 
the suppression of pion exchange reactions such as $e d  \to e + {\rm forward} \,\, \Delta^-  + {\rm slow} \,\,  
\Delta^{++}$~\cite{Frankfurt:1996ai}.

\section {QCD and the Internucleon Interaction}

 The chiral QCD dynamics modified by including $\Delta$ isobars is used  in the calculation of two-pion exchange 
potential, as can be 
 found, e.g., in~\cite{Krewald:2012zza}. This is an example of how the inner structure of a nucleon enters nuclear theory. 
The lack of collapse of  heavy nuclei and the form of the dominance of the repulsion in $NN $ interaction at short distances  
in the phenomenological potential of $NN$ interaction suggests the 
important role of strong internucleon repulsion at the distances $r_{NN}\le 0.5$~fm.  
The observation of the high momentum tail of the nucleon momentum  distribution allows to restrict the form of 
internucleon interaction~\cite{Frankfurt:2008zv,Hen:2016kwk}.

  A pion is an not elementary particle but  a pseudo-Goldstone boson of QCD. 
With the nuclear density increase, an interacting  nucleon loses  the pion cloud because the pQCD phase has no pseudo-Goldstone bosons. 
Thus, the chiral symmetry restoration phase transition starts  at the nuclear  densities as a factor of 
$\ge (1.4\, {\rm fm}/2r_c)^3$, approximately three times  
larger than  the saturation nuclear~density.

Intermediate range internucleon forces are due to exchange by constituent quarks between nucleons.  This is because the 
chiral condensate which gives mass to light  quarks is mostly due to the instanton field of a quark;
cf. Ref.~\cite{Ioffe:2005ym}.

 The pQCD core within a nucleon has important implications for the theory of super-dense nuclear matter, i.e., for the 
inner cores of neutron stars. At the nucleon densities where the internucleon repulsion dominates,  $r_{NN}\le 0.5$ --  
0.4 fm, 
the pressure from adjacent nucleons leads to the squeezing of a nucleon, i.e., to increase 
 $x_i$ and 
$k_{i t}$
 compared to 
 those 
at the saturation of nuclear density.  At these densities, only the pQCD cores of nucleons survives. If the density  
further increases to 
 (1.4~fm/(0.5--0.7)~fm)$^3$, 
approximately 8 to 22 
times the  nuclear density, the chiral restoration phase transition 
 should take place. This is obviously a rather 
rough estimate, which 
is given
here 
merely for 
illustration.

 The interaction between a quark--gluon core and another nucleon becomes repulsive with the increase in nuclear density. When the 
internucleon distances itself 
from $r_{NN}\ll 2r_c\approx 1 fm$, the interaction becomes 
 of strong repulsion. 

It was suggested~\cite{Zeldovich:1974zn} that the  repulsion is mainly due to the  Pauli principle for quarks belonging to the different nucleons.

\section{Conclusions}

 We argued that a nucleon in QCD (quantum chromodynamics) 
contains three coexisting layers of two QCD phases. An external layer of 
a nucleon is formed by the pion cloud of a nucleon and dominates in the low-energy nuclear
phenomena in the form of the attractive potential of 
 nucleon--nucleon 
interaction. The perturbative QCD (pQCD)
core of a nucleon leads to the existence of a large group of hard
processes off  nuclear targets, where the onset of  the color transparency (CT)} phenomena is rapid, as described within the
concept of QCD factorization theorems. The nonperturbative phase
of spontaneously broken chiral symmetry surrounding the pQCD core of a
nucleon reveals itself in the slow onset of a CT  phenomenon for
quasi-elastic processes off  nuclear targets. The account of the Sudakov form
factors accelerates  the onset of CT but only slightly. Thus, the three -layer
structure of a nucleon leads to the dominance of hadron degrees of
freedom at the average internucleon distances in  nuclei, at the
saturation nuclear density,  and to the restoration of the chiral symmetry with
a  further increase in the nuclear density.

Systematic experimental studies of possible CT effects in two-body processes should continue employing 
the projectiles and produced hadrons of  different sizes (direct photons~\cite{Larionov}, pions, kaons, etc.).  One should also  
study  $2\to 3$ high-energy processes such as $\pi^- + A\to \pi^-\pi^- (A-1)^* $ with back-to-back pions  with 
 transverse momentum,
$p_t  > 1.5$ 
 \hl{to}  
2 GeV/$c$~\cite{Larionov} which could be measured in the COMPASS experiment at CERN. In this 
kinematics, the projectile pion  and two final pions remain frozen over distances exceeding by far the size \hl{of} 
heavy nucleus, namely, with the coherence length, 
$L_c\sim $ 60 fm~\cite{Kumano}.
\vspace{6pt}

\authorcontributions{The authors contributed equally to this work. All authors have read and agreed to the published version of the manuscript.

\funding{This research was supported by the U.S. Department of Energy  grant DE-FG02-93ER40771  and by the Binational Science Foundation United States--Israel grant 202115.}}

\acknowledgments{ We thank Misak Sargsyan for organizing the volume.}


\conflictsofinterest{Authors declare no conflict of interest.}


\begin{adjustwidth}{-\extralength}{0cm}

\reftitle{References}

\end{adjustwidth}

\end{document}